\documentclass[showpacs,amsmath,amssymb,aps]{revtex4}

\usepackage{graphicx}
\usepackage{dcolumn}
\usepackage{bm}
\usepackage{epsfig}

\begin{document}

\title{The baryon mass calculation in the chiral soliton model at finite temperature and density}
\author{Hui~Zhang, Renda~Dong}
\author{Song~Shu}\email[Corresponding author.\\E-mail address:]{shus@hubu.edu.cn}
\affiliation{Department of Physics and Electronic Technology, Hubei University, Wuhan 430062, China}

\begin{abstract}
In the mean-field approximation, we have studied the soliton which is embedded in a thermal medium within the chiral soliton model. The energy of the soliton or the baryon mass in the thermal medium has been carefully evaluated, in which we emphasize that the thermal effective potential in the soliton energy should be properly treated in order to derive a finite and well-defined baryon mass out of the thermal background. The result of the baryon mass at finite temperatures and densities in chiral soliton model are clearly presented.
\end{abstract}
\maketitle

\section{Introduction}

Quantum chromodynamics (QCD), as the fundamental theory of strong interactions, gains an increasing number of applications in particle physics. Perturbative QCD calculations (lattice QCD) methods work well for small distances or high energies. However, the analytical as well as the numerical methods have not been developed enough on the scale of large distances or low energies due to its essential non-perturbative features of hidden (spontaneously broken) chiral symmetry and confinement, especially if the baryons are involved \cite{ref1,ref2,ref3,ref4,ref5,ref6,ref7}. Therefore, effective quark models which have the certain essential features of QCD are other important choices \cite{ref8}. The chiral soliton model which is also called the linear sigma model (L$\sigma$M) \cite{ref9}, as one of effective models, incorporates  the chiral symmetry and its spontaneous breaking \cite{ref10,ref11,ref12,ref13,ref14}. The model which has been proposed as a model for strong nuclear interactions provides a good description of the nuclear properties. The chiral soliton model could be solved in the mean-field approximation. A semiclassical soliton solution is referred to a baryon, and the baryon properties can be derived easily from the soliton solution \cite{ref15,ref16,ref17,ref18,ref19,ref20}. 

In recent years, studies on the behavior of strongly interacting mater under extreme conditions which is created by relativistic heavy ion collision are more and more interesting. The hadron properties (masses, radius, magnetic moments, etc.) in a hot and dense medium have drawn a lot of attentions \cite{ref21,ref22,ref23}. The hadron properties reflect the non-perturbative features of QCD. At zero temperature, there are sufficient studies on the hadron properties, especially the hadron mass, within effective models. At finite temperature, the quarks (and thus hadrons) are expected to become lighter with chiral symmetry getting restored in the famous Brown-Rho paper \cite{ref24}. This meets quite many papers, such as references \cite{ref33,ref34,ref35}. However, there are also many papers which obtain that the hadron mass increase with the temperature increasing \cite{ref28,ref29,ref30}.

The chiral soliton model successfully predicts the static nucleon properties at zero temperature and density \cite{ref15}. The nucleon properties has also been studied through the same model at finite temperatures and densities \cite{ref25,ref26,ref27,ref28,ref29,ref30}. The effective masses have been obtained by these solitons. In these studies, however, the thermal medium contribution to the soliton energy is not properly treated. In reference \cite{ref25,ref26,ref27}, the energy of the thermal medium is completely neglected, but under this treatment the baryon mass could not go back to the right result of the baryon mass at zero temperature and density. In references \cite{ref28,ref29,ref30}, the thermal medium contribution is not well subtracted which results in unphysical rising of a baryon mass at high temperatures. Our goal in this paper is to give a well defined baryon mass calculation in a thermal medium within the soliton model.

The structure of this paper is as follows: In section \ref{sec:model}, the chiral soliton model is introduced at zero as well as finite temperature and density. In section \ref{sec:mass}, the baryon mass calculation in medium through chiral soliton model is discussed. In section \ref{sec:results}, we present the thermal effective potential density, and show the soliton solutions of the chiral soliton equations at different temperatures and densities. Then the numerical results are discussed before the summary section.

\section{The chiral soliton model at finite temperature and density}\label{sec:model}

The Lagrangian density of the chiral soliton model with the interactions of quarks and mesons is \cite{ref15}
\begin{equation}
\mathcal L = \bar\psi [ i \gamma_\mu \partial^\mu + g (\sigma + i \gamma_5 \vec\tau \cdot \vec\pi )] \psi + \frac{1}{2} ( \partial_\mu \sigma \partial^\mu \sigma + \partial_\mu \vec\pi \partial^\mu \vec\pi ) - U ( \sigma , \vec{\pi} ) ,\label{L}
\end{equation}
where $\psi$ represents the spin-$\frac{1}{2}$ two flavors light quark fields $\psi=(u,d)$, $\sigma$ is the spin-0 isosinglet scalar field, and $\vec\pi$ is the spin-0 isovector pion field $\vec\pi=(\pi_1,\pi_2,\pi_3 )$. The potential for $\sigma$ and $\vec\pi$ is
\begin{equation}
U ( \sigma , \vec\pi ) = \frac{\lambda}{4} ( \sigma^2 + \vec\pi^2 - \nu^2 )^2 + H \sigma - \frac{m_\pi^4}{4 \lambda} + f_\pi^2 m_\pi^2 ,\label{U}
\end{equation}
where the last two constant terms in equation (\ref{U}) are used to guarantee that the energy of a vacuum in the absence of quarks is zero. The minimum energy occurs for chiral fields $\sigma$ and $\vec\pi$ restricted to the chiral circle
\begin{equation}
\sigma^2 + \vec\pi^2 = f_\pi^2,
\end{equation}
where $f_\pi=93 MeV$ is the pion decay constant, $H \sigma$ is the explicit chiral symmetry breaking term, $H=f_\pi m_\pi^2$, and $m_\pi=138 MeV$ being the pion mass. The chiral symmetry is explicitly broken in vacuum and the expectation values of the meson fields are: $\langle\sigma\rangle=-f_\pi$ and $\langle\vec\pi\rangle=0$. The constituent quark mass in vacuum is $M_q=g f_\pi$, and the $\sigma$ mass is defined by $m_\sigma^2=m_\pi^2+2\lambda f_\pi^2$. The quantity $\nu^2$ can be expressed as $\nu^2=f_\pi^2-m_\pi^2/\lambda$. In our calculation we follow the choice of the reference~\cite{ref15} and set the constituent quark mass and the sigma mass as $M_q=500 MeV$ and $m_\sigma=1200 MeV$ that determine the parameters $g \approx 5.28$ and $\lambda \approx 82.1$.\\

In order to investigate the temperature and chemical potential dependence of the chiral soliton, we embed a single soliton in a homogeneous hot and dense quark medium with temperature $T$ and chemical potential $\mu$. First, we derive the thermal effective potential of the spatially uniform system at finite temperature and density using the finite temperature field theory \cite{ref31}.
\begin{equation}
\Omega (\sigma,\pi;T,\mu) = U(\sigma,\pi ) + \Omega_{\bar{\psi} \psi}+ B(T,\mu) ,  \label{Omega}
\end{equation}
where $B(T,\mu)$ is used to guarantee that the absolute minimum value of the thermal effective potential is zero, and it is the key of strictly calculating the baryon mass, which will be discussed later. The thermodynamical potential which is distributed by the homogeneous medium is
\begin{equation}
\Omega_{\bar{\psi} \psi} = -\nu_q T \int \frac{\mathrm{d^3} \vec{p}}{(2\pi )^3} \{ \ln[ 1+ e^{-(E_q - \mu)/T}] + \ln[ 1+ e^{-(E_q + \mu)/T}] \} ,
\end{equation}
where $\nu_q$ is the degeneracy factor $\nu_q=2(spin) \times 2(flavor) \times 3(color)=12$ and $E_q=\sqrt{\vec p^2+M_q^2}$ is the valence quark and antiquark energy for $u$,$d$ quarks. The constituent quark (antiquark) mass $M_q$ is defined by
\begin{equation}
M_q^2 = g^2 (\sigma^2 + \pi^2) .
\end{equation}
At large radius r, the $\sigma$ field assumes its vacuum value $\sigma_v$. The minimum energy either in a vacuum or in a thermal vacuum for chiral fields is restricted to the chiral circle
\begin{equation}
\sigma^2 + \pi^2= \sigma_v^2 ,
\end{equation}
where the value of $\sigma_v$ in the thermal medium should be determined by the absolute minimum of the thermodynamical potential, which is $\frac{\partial \Omega }{\partial \sigma}=0$ \cite{ref32}.

Now we embed a soliton in a homogeneous hot and dense quark medium with temperature $T$ and chemical potential $\mu$. Thus the effective Lagrangian is
\begin{equation}
\mathcal L_{eff} = \bar\psi [ i \gamma_\mu \partial^\mu + g (\sigma + i \gamma_5 \vec\tau \cdot \vec{\pi} )] \psi + \frac{1}{2} ( \partial_\mu \sigma \partial^\mu \sigma + \partial_\mu \vec{\pi} \partial^\mu \vec{\pi} ) -\Omega (\sigma,\pi;T,\mu) ,\label{Leff}
\end{equation}
The lagrangian can also be found in references~\cite{ref25,ref26,ref27,ref28}. From the Lagrangian, the field radial equations  at finite temperature and density could be derived
\begin{eqnarray}
\frac{\mathrm{d} u(r)}{\mathrm{d} r} = -( \epsilon - g \sigma (r)) v(r) - g \pi(r) u(r) ,\label{u}\\
\frac{\mathrm{d} v(r)}{\mathrm{d} r} = -( \frac{2}{r} - g \pi(r)) v(r) + ( \epsilon + g \sigma(r)) u(r) ,\label{v}\\
\frac{\mathrm{d}^2 \sigma (r)}{\mathrm{d} r^2}  + \frac{2}{r} \frac{\mathrm{d} \sigma(r)}{\mathrm{d} r} + N g (u^2(r) - v^2(r)) = \frac{\partial \Omega}{\partial \sigma} ,\label{sigma}\\
\frac{\mathrm{d}^2 \pi(r)}{\mathrm{d} r^2} + \frac{2}{r} \frac{\mathrm{d} \pi(r)}{\mathrm{d} r} - \frac{2\pi(r)}{r^2} + 2 N g u(r) v(r) = \frac{\partial \Omega}{\partial \pi} .\label{pi}
\end{eqnarray}
where
\begin{eqnarray}
\frac{\partial \Omega}{\partial \sigma} = \frac{\partial U(\sigma,\pi)}{\partial \sigma} + g^2 \sigma \nu_q \int \frac{\mathrm{d^3} p}{(2 \pi)^3} \frac{1}{E_q} (\frac{1}{1+ e^{(E_q - \mu)/T}} + \frac{1}{1+ e^{(E_q + \mu)/T}}) ,\label{sigma2}\\
\frac{\partial \Omega}{\partial \pi} = \frac{\partial U(\sigma,\pi)}{\partial \pi} + g^2 \vec{\pi} \nu_q \int \frac{\mathrm{d^3} p}{(2 \pi)^3}\frac{1}{E_q} (\frac{1}{1+ e^{(E_q - \mu)/T}} + \frac{1}{1+ e^{(E_q + \mu )/T}}) .\label{pi2}
\end{eqnarray}

In the above derivations one takes the mean-field approximation and the ``hedgehog" ansatz witch means
\begin{eqnarray}
&& \langle \sigma (\vec{r},t) \rangle = \sigma (r)  ,\ \ \ \ \langle \vec{\pi} (\vec{r},t) \rangle = {\vec{r}} \pi(r) ,\\
&& \psi(\vec{r},t) = e^{-i \epsilon t} \sum_{i=1}^{N} q_i(\vec{r}),\ \ \ \ q(\vec{r}) = \binom{u(r)}{i \vec{\sigma} \cdot {\vec{r}} v(r)} \chi ,\\
&& (\vec{\sigma} + \vec{\tau}) \chi = 0,
\end{eqnarray}
where $q_i$ are N identical valence quarks in the lowest s-wave level with (eigen) energy $\epsilon$. $N$ is set to 3 for baryons and 2 for mesons. $\chi$ is the spinor. The quark functions should satisfy the normalization condition
\begin{equation}
4\pi \int r^2(u^2(r) + v^2(r)) \mathrm{d}r = 1 \label{n}.
\end{equation}

And the boundary conditions are
\begin{eqnarray}
v(0)=0 ,\ \ \ \frac{\mathrm{d} \sigma (0)}{\mathrm{d} r}=0 ,\ \ \ \pi(0)=0 , \label{b1}\\
u(\infty)=0 ,\ \ \ \sigma(\infty)=\sigma_v ,\ \ \ \pi (\infty )=0 . \label{b2}
\end{eqnarray}

\section{The baryon mass calculation in medium}\label{sec:mass}

At certain values of temperature and density, the equations (\ref{u})-(\ref{pi}) together with normalization condition (\ref{n}) and boundary conditions (\ref{b1}),(\ref{b2}) which are nonlinear ordinary differential equations could be numerically solved. Using this solution, the physical properties of the three-quark system can be calculated. The total energy or mass of the hedgehog baryon is given by:
\begin{eqnarray}
E = M_B = N \epsilon + 4 \pi \int_0^\infty {\mathrm{d} r} \mathcal{E},   \label{M}
\end{eqnarray}
where N is set to 3 for baryon and $\epsilon$ is the quark energy. At zero temperature and density its value is 30.5MeV. When the soliton equations are solved at finite temperatures and densities, it will change with the temperature and density, which will be presented in the next section.

$\mathcal{E}(r;T,\mu)$ in equation (\ref{M}) is radial energy density of the meson fields, and it reads
\begin{eqnarray}
\mathcal{E}=r^2 [\frac{1}{2}(\frac{\mathrm{d} \sigma}{\mathrm{d} r})^2 + \frac{1}{2}(\frac{\mathrm{d} \pi}{\mathrm{d} r})^2 + \frac{\pi ^2}{r^2} + \Omega(\sigma,\pi;T,\mu)] , \label{E}
\end{eqnarray}
where $\Omega$ is the thermal effective potential. This potential energy plays dual roles in the system. Assume $r_0$ as the radius of a soliton. Inside the domain of the soliton or $r\sim r_0$, $\Omega$ is the effective potential energy of the meson fields. While outside the domain of the soliton or $r\gg r_0$, $\Omega$ is the thermodynamic potential energy of the homogeneous medium. Since it includes the energy of the thermal background, it should be properly subtracted off. Otherwise the integral of the soliton energy would be infinite. In some previous studies \cite{ref25,ref26,ref27} this energy has been completely neglected, in order to make the integral finite. However, this treatment is not proper, as the soliton energy could not go back to the right form when temperature and density go to zero. In other studies  \cite{ref28,ref29,ref30}, the background energy has not been well subtracted, and the baryon mass becomes unphysically large at high temperatures or densities as a result.

Now, let us make an analysis of the integral of the soliton energy. At zero temperature and density, the thermal effective potential $\Omega$ becomes to the potential $U(\sigma,\pi)$. When $r\rightarrow \infty$, the meson fields $\sigma$ and $\pi$ assume their vacuum values, which are determined by $\frac{\partial U(\sigma,\pi)}{\partial \sigma}=\frac{\partial U(\sigma,\pi)}{\partial \pi}=0$. At these values, the potential $U(\sigma,\pi)$ has the lowest minimum energy which is zero corresponding to the vacuum. From (\ref{M}) one could see that the integral is finite as $r\rightarrow \infty$, $U(\sigma,\pi)\rightarrow 0$. Now let us take a look of the case at finite temperature and densitiy. $U(\sigma,\pi)$ is replaced by $\Omega(\sigma,\pi)$. When $r\rightarrow \infty$, the meson fields $\sigma$ and $\pi$ assume their thermal vacuum values, which are determined by $\frac{\partial \Omega(\sigma,\pi)}{\partial \sigma}=\frac{\partial \Omega(\sigma,\pi)}{\partial \pi}=0$. At these values, the thermal effective potential $\Omega(\sigma,\pi)$ has the lowest minimum energy but nonzero which represents the energy of the thermal vacuum. From (\ref{M}) one could see that the integral is infinite because $\Omega(\sigma,\pi)$ approaches a nonzero value as $r\rightarrow \infty$. The thermal background energy has been included in evaluating the soliton energy, therefore it is infinite. 

How to properly subtracted this background energy? We think that one should make a redefinition of the energy of the thermal vacuum when evaluating the integral, that is to say, when $r\rightarrow \infty$, one should set $\Omega(\sigma,\pi)\rightarrow 0$. This could be fulfilled by readjust the $B(T,\mu)$ to make the minimum of thermodynamic potential $\Omega$ always staying at zero at different temperatures and densities. By this treatment, the integral becomes finite, and the energy of the thermal background has been successfully subtracted off. Thus we obtain a finite and well defined baryon mass in thermal medium.

\section{Numerical results}\label{sec:results}

\begin{figure}[tbh]
\includegraphics[width=210pt,height=150pt]{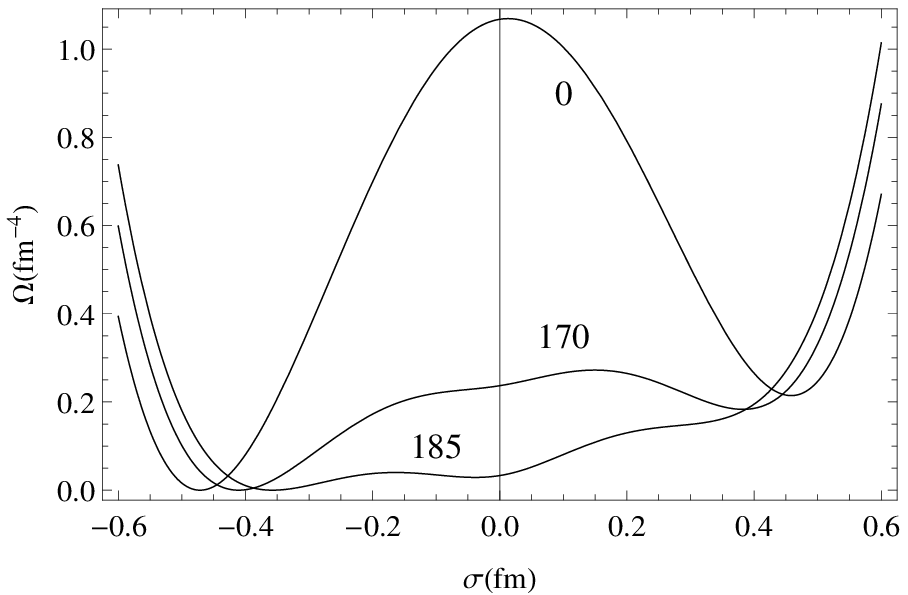}
\hspace{1cm}
\includegraphics[width=210pt,height=150pt]{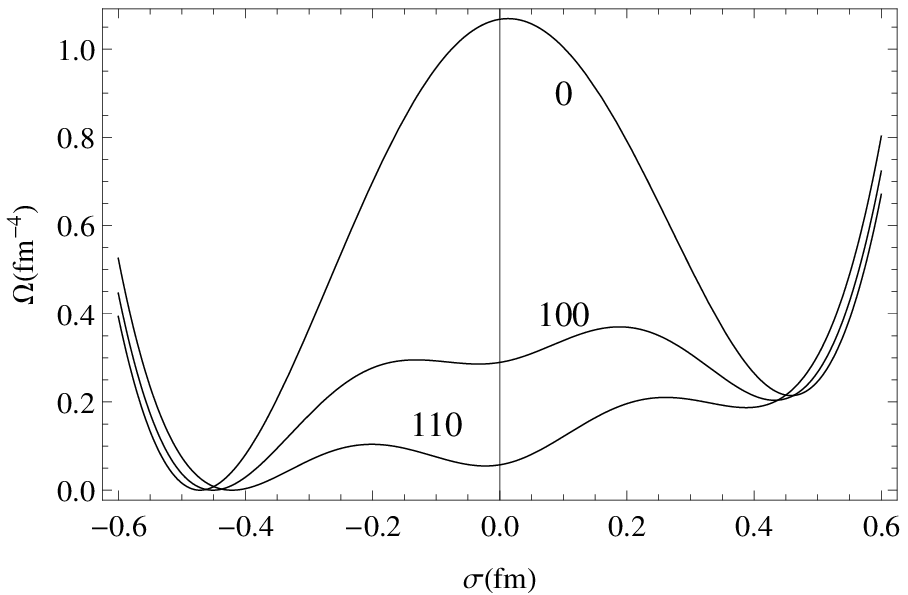}\\
(a)\; \; \; \; \; \; \; \; \; \; \; \; \; \; \; \; \; \; \; \; \; \; \; \; \; \; \; \; \; \; \; \; \; \; \; \; \; \;  (b)
\caption{The thermodynamical potential $\Omega$, (a) at $\mu=0$, $T=\{0,170,185\} MeV$. (b) at $\mu=300 MeV$, $T=\{0,100,110\} MeV$.} \label{f1}
\end{figure}

It is instructive to plot the thermal effective potential as a function of the $\sigma$ filed for different temperatures and densities. In Fig.\ref{f1}, the left part (a) is for $\mu=0$ and the right (b) for $\mu=300 MeV$. One can see that by adding the temperature and density dependent parameter $B(T,\mu)$ we have shifted the minimum value of thermal effective potential to zero for different temperatures and densities. This means when $\sigma=\sigma_v$ at different temperatures and densities we always have $\Omega=0$. As a result, the thermal background energy has been subtracted out of the soliton energy as in (\ref{M}). In our subtraction approach, $B(T,\mu)$ plays an important role in deriving the finite baryon mass. At different temperatures and densities, the numerical results of $B(T,\mu)$ are shown in Table.\ref{t1}. It is sensitive to the variation of the temperature, but not of the chemical potential. $B(T,\mu)$ increases with the temperature and chemical potential increasing.

\begin{figure}[tbh]
\includegraphics[width=210pt,height=150pt]{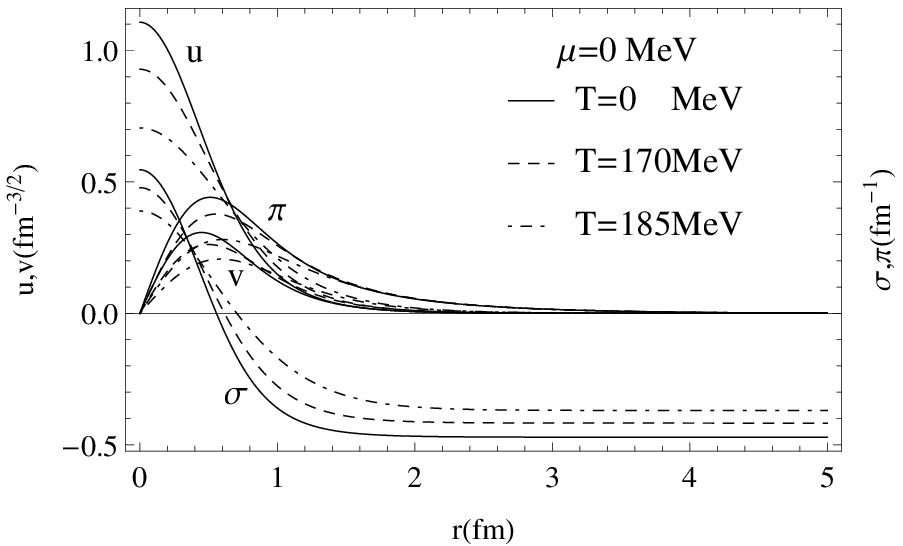}
\hspace{1cm}
\includegraphics[width=210pt,height=150pt]{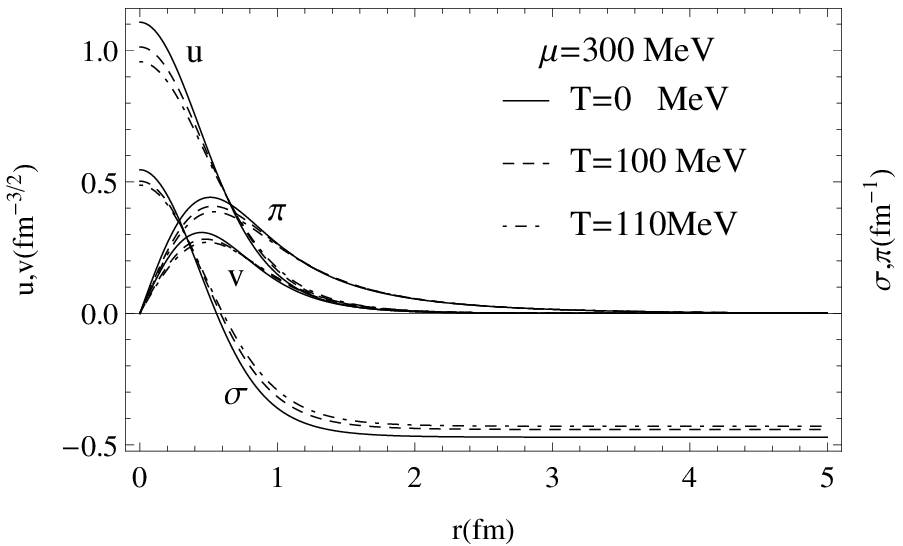}\\
(a)\; \; \; \; \; \; \; \; \; \; \; \; \; \; \; \; \; \; \; \; \; \; \; \; \; \; \; \; \; \; \; \; \; \; \; \; \; \; \; \; (b)
\caption{The quark fields $u(r)$, $v(r)$ and the meson fields $\sigma(r)$, $\pi(r)$ as functions of the radius $r$, (a) at $\mu=0 MeV$, $T=\{0,170,185\} MeV$. (b) at $\mu=300 MeV$, $T=\{0,100,110\} MeV$.}\label{f2}
\end{figure}
\begin{table}[tbh]
\caption{The energy or mass $M_B$ of baryon, eigenvalue $\epsilon$ and $B(T,\mu)$  at different temperatures and chemical potentials.}
\begin{ruledtabular}
\begin{tabular}{ccccc}
$\mu=0$  &  T(MeV)  &  0  &  170  &  185   \\
\hline
&  $\epsilon(fm^{-1})$   & 0.216  & 0.289  & 0.451  \\
&  $B(fm^{-4})$   & 0  & 0.439  & 0.746  \\
\hline\hline
$\mu=300$MeV  &  T(MeV)  &  0  &  100  &  110   \\
\hline
&  $\epsilon(fm^{-1})$   & 0.216  & 0.257  & 0.288  \\
&  $B(fm^{-4})$   & 0  & 0.074  & 0.202  \\
\end{tabular}
\end{ruledtabular} \label{t1}
\end{table}

In Fig.\ref{f2}, we plot the quark fields $u(r)$, $v(r)$ and the meson fields $\sigma(r)$, $\pi(r)$ as functions of the radius $r$ at fixed chemical potential for different temperatures: the left (a) is for $\mu=0$ and the right (b) for $\mu=300 MeV$. It can be seen that in both cases the amplitudes of the soliton solutions decrease and change more and more rapidly with the temperature increasing.

In solving $u$ and $v$ fields, one should notice that the normalization condition (\ref{n}) must be observed at different temperature and densities. This makes quark energy $\epsilon$ changing with temperature and density, which is shown in Table.\ref{t1}. $\epsilon$ increases with the temperature and chemical potential increasing. From $\sigma$ and $\pi$ together with thermal effective potential $\Omega$ in which the thermal background energy has been subtracted off, one can obtain the energy density $\mathcal{E}$ by (\ref{E}).

\begin{figure}[tbh]
\includegraphics[width=210pt,height=150pt]{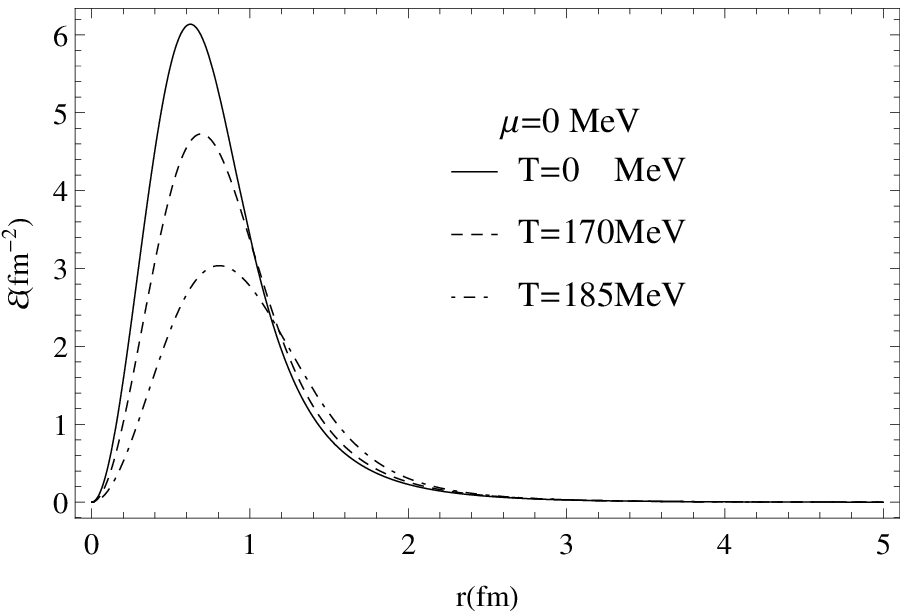}
\hspace{1cm}
\includegraphics[width=210pt,height=150pt]{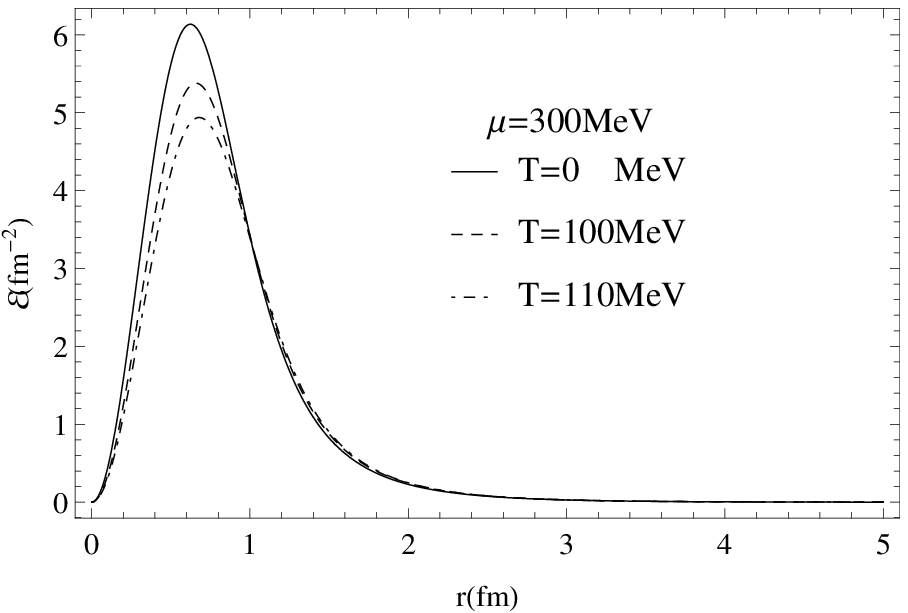}\\
(a)\; \; \; \; \; \; \; \; \; \; \; \; \; \; \; \; \; \; \; \; \; \; \; \; \; \; \; \; \; \; \; \; \; \; \; \; \; \; \; \; (b)
\caption{The energy density $\mathcal{E}(r)$ as functions of the radius $r$, (a) at $\mu=0$, $T=\{0,170,185\} MeV$. (b) at $\mu=300 MeV$, $T=\{0,100,110\} MeV$. } \label{f3}
\end{figure}

In Fig.\ref{f3}, the energy density $\mathcal{E}$ are plotted as functions of $r$ at different temperatures for $\mu=0$ and $\mu=300 MeV$. One can see that as $r\rightarrow \infty$, we have $\mathcal{E} \rightarrow 0$. This can only be fulfilled when the thermal background energy is well subtracted off. One can also see that the amplitude of $\mathcal{E}$ decreases more and more rapidly, and the width of $\mathcal{E}$ gets ``fatter" with the temperature increasing. The position of the peak value which represents the radius of the soliton increases with the temperature increasing. It means that the spatially localized energy distribution expands with the temperature increasing. One can see from Fig.\ref{f3}, the area under the curve of function $\mathcal{E}(r)$ diminishes with the temperature increasing, which results in that the meson energy decreases with the temperature increasing. 

\begin{figure}[tbh]
\includegraphics[width=210pt,height=150pt]{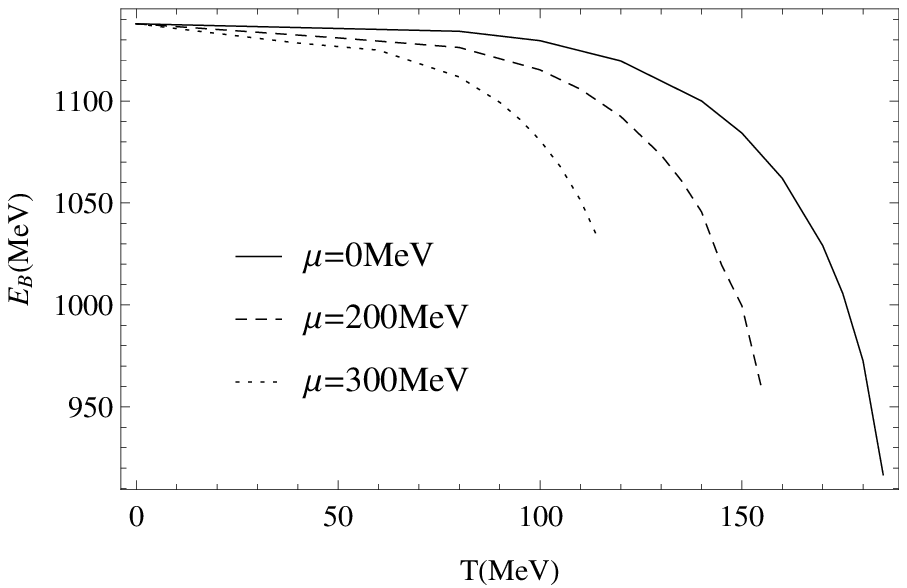}
\hspace{1cm}
\includegraphics[width=210pt,height=150pt]{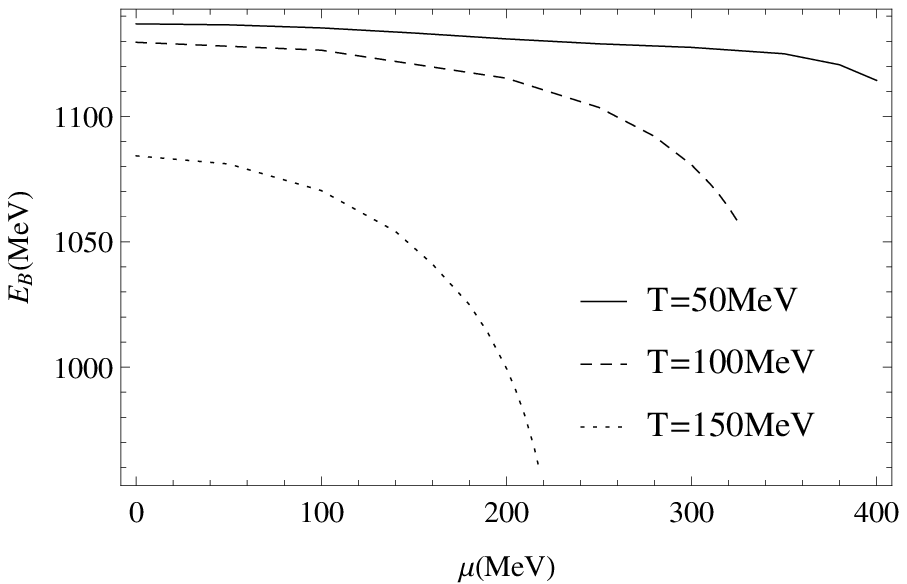}\\
(a)\; \; \; \; \; \; \; \; \; \; \; \; \; \; \; \; \; \; \; \; \; \; \; \; \; \; \; \; \; \; \; \; \; \; \; \; \; \; \; \; (b)
\caption{The Baryon mass $E_B$ of a stable chiral soliton (a) as a function of the temperature $T$ at $\mu=0$, $\mu=200 MeV$ and $\mu=300 MeV$. (b) as a function of the temperature $\mu$ at $T=50 MeV$, $T=100 MeV$ and $T=150 MeV$. }\label{f4}
\end{figure}

From Eq.(\ref{M},\ref{E}), one could see that the soliton energy comes from the summation of the quark energy and the meson energy. Although the quark energy is increasing with temperature or chemical potential increasing, the decrease of the meson energy outweighs the increase of quark energy. As a result the total energy or the baryon mass will decrease with temperature or chemical potential increasing, which is presented in Fig.\ref{f4}.  And the baryon mass $E_B$ decreases more and more rapidly with the temperature or chemical potential increasing. Other works employing the same model had presented the baryon mass \cite{ref28,ref29,ref30,ref25}. Our curve is quite similar with the curve $E^*$ of Fig.4 in Ref.\cite{ref25}. However, they treated the effective potential in the soliton energy as the background medium attribution, and completely neglected it, which also makes the baryon mass finite. From our discussion here, it could be seen this treatment of subtracting the background energy is not proper. In their results, the rate of decline of baryon mass with temperature increasing is larger than ours. In references \cite{ref28,ref29,ref30}, the authors had not subtracted the thermal background energy from the thermodynamic potential. Therefore, in their results the baryon mass is infinite, but they had just made a cut-off. This scheme made the baryon mass unphysically large at high temperatures or densities. In references \cite{ref26,ref27}, the authors had obtained the nucleon properties through the soliton in the NJL model. The nucleon mass decreases with temperature increasing at zero density, while at finite density it increases at first and then decreases with temperature increasing. At high temperatures, the qualitative results of the decreasing of the nucleon mass is consistent with ours and those in Ref.\cite{ref25}.

\section{Summary}

In this paper, we have studied the chiral soliton model at finite temperature and density, and solved the chiral soliton equations at different temperatures and densities with different boundary conditions. By properly subtracting the thermal background energy we have obtained a strictly well-defined finite baryon mass in soliton model at finite temperature and density. As a result the baryon mass decreases with the temperature and chemical potential increasing, which is consistent with the Brown-Rho scaling.

\begin{acknowledgments}
This work was supported in part by the National Natural Science Foundation of China with No. 10905018 and No. 11275082.
\end{acknowledgments}

\end{document}